\documentstyle[12pt]{article}
\def\singlespace {\smallskipamount=3.75pt plus1pt minus1pt
                  \medskipamount=7.5pt plus2pt minus2pt
                  \bigskipamount=15pt plus4pt minus4pt
                  \normalbaselineskip=15pt plus0pt minus0pt
                  \normallineskip=1pt
                  \normallineskiplimit=0pt
                  \jot=3.75pt
                  {\def\smallskip {\vskip\smallskipamount}}
                  {\def\medskip   {\vskip\medskipamount}}
                  {\def\bigskip   {\vskip\bigskipamount}}
                  {\setbox\strutbox=\hbox{\vrule
                    height10.5pt depth4.5pt width 0pt}}
                  \parskip 7.5pt
                  \normalbaselines}
\def\middlespace {\smallskipamount=5.625pt plus1.5pt minus1.5pt
                  \medskipamount=11.25pt plus3pt minus3pt
                  \bigskipamount=22.5pt plus6pt minus6pt
                  \normalbaselineskip=22.5pt plus0pt minus0pt
                  \normallineskip=1pt
                  \normallineskiplimit=0pt
                  \jot=5.625pt
                  {\def\smallskip {\vskip\smallskipamount}}
                  {\def\medskip   {\vskip\medskipamount}}
                  {\def\bigskip   {\vskip\bigskipamount}}
                  {\setbox\strutbox=\hbox{\vrule
                    height15.75pt depth6.75pt width 0pt}}
                  \parskip 11.25pt
                  \normalbaselines}
\def\doublespace {\smallskipamount=7.5pt plus2pt minus2pt
                  \medskipamount=15pt plus4pt minus4pt
                  \bigskipamount=30pt plus8pt minus8pt
                  \normalbaselineskip=30pt plus0pt minus0pt
                  \normallineskip=2pt
                  \normallineskiplimit=0pt
                  \jot=7.5pt
                  {\def\smallskip {\vskip\smallskipamount}}
                  {\def\medskip   {\vskip\medskipamount}}
                  {\def\bigskip   {\vskip\bigskipamount}}
                  {\setbox\strutbox=\hbox{\vrule
                    height21.0pt depth9.0pt width 0pt}}
                  \parskip 15.0pt
                  \normalbaselines}
\newcommand{\beqn}{\begin{equation}}
\newcommand{\eeqn}{\end{equation}}
\begin{document}
\vspace*{-1in}
\begin{flushright}
CUPP-96/2 \\
TIFR/TH/96-18\\[10pt]
hep-ph/9604285n
\end{flushright}
\vskip 40pt
\begin{center}

{\large{\bf Unification and Model Builiding,
\vskip 5pt
Astroparticle Physics and Neutrinos}}
\vskip 10pt
{\bf WHEPP4 Working Group Report}
\vskip 30pt
Amitava Raychaudhuri\\
{\it Department of Pure Physics, University of Calcutta, \\
92 Acharya Prafulla Chandra Road, Calcutta - 700 009, India}\\

\vskip 10pt
and
\vskip 10pt

Probir Roy\\
{\it Tata Institute of Fundamental Research, Homi Bhabha Road,\\
Bombay - 400 005, India}\\
 
\vskip 30pt

Coordinators \\

\vskip 10pt

R. Gandhi, A. Joshipura, P.B. Pal, A.  Raychaudhuri, P. Roy, U.
Sarkar\\

\vskip 10pt

Participants \\

\end{center}

R. Adhikari, R. Barbieri, G. Bhattacharyya, B. Brahmachari, D.
Choudhury, S. Dutta, R. Gandhi, A. Ghosal, S. Goswami, A. Joshipura, K.
Kar, Amit Kundu, U.  Mahanta, S. Mallik, Mamta, A. Mishra, H. Mishra,
B. Mukhopadhyaya, H.  Murayama, H. Nagar, S. Nandi, S. Pakvasa, P.B.
Pal, M.K. Parida, G.  Rajasekaran, M.V. Ramana, A.K. Ray, A.
Raychaudhuri, M. Roy, P. Roy, M.K. Samal, S. Sarkar, U.
Sarkar, V. Sheel

\vskip 50pt

\begin{center}
{\bf Abstract}
\end{center}

This report summarises the work done during the Workshop on High Energy
Physics Phenomenology 4 (S.N. Bose National Centre for Basic Sciences,
Calcutta, India, Jan 2-14, 1996) in Working 
Groups IV (Unification and Model Building) and V (Astroparticle Physics
and Neutrinos).

\vskip 30pt
April 1996

\newpage
 
 
\section {Introduction}

The original plan was to have two separate Working
Groups called `Unification and Model Building' and `Astroparticle Physics
and Neutrinos'.  However, at the beginning of the Workshop, it
emerged that the members of these two working groups had a wide overlap
of interest. Hence a decision was taken to merge them into one working
group. 
In this working group several problems were initially identified for
investigation. Smaller subgroups were formed to focus on each of these
problems. The progress till the end of the workshop is summarised in
this report.

\section {Problems undertaken}

\begin{enumerate}

\item {\bf CP odd $WW\gamma$ and $WWZ$ form-factors in the MSSM}

Possible nonstandard couplings in the $WW\gamma$ and $WWZ$ vertices
and how to probe them at the Next Linear Collider (NLC) have been
well-studied in the CP-conserving sector.  In principle, however,
nonstandard interactions could lead to anomalously large CP-odd
couplings too.  These had not been studied so far.
The CP-odd part of the $W-W-V$ vertex $(V = \gamma,Z)$ can be
expressed in terms of the 
two couplings $\widetilde{\kappa_V}$ and $\widetilde{\lambda_V}$ {\em
via}
\beqn
{\cal L}_{eff}^{WWV} = \frac{i}{2} \left[ 
\widetilde{\kappa_V} W^+_{\mu} W^-_{\nu} \epsilon^{\mu \nu \rho \sigma}
V_{\rho \sigma} + \frac{\widetilde{\lambda_V}}{m_W^2} 
W^+_{\alpha \mu} W^{- \alpha}_{\nu} \epsilon^{\mu \nu \rho \sigma}
V_{\rho \sigma} \right]. 
\eeqn
At the NLC $\widetilde{\kappa_V}$ and
$\widetilde{\lambda_V}$ may be probed at the level of $10^{-3}$. Within
the Standard Model (SM) the one loop contributions to these couplings
vanish. Such is not the case for the Minimal Supersymmetric Standard
Model (MSSM). It is found that

\begin{itemize}

\item Triangle diagrams with scalars in the loop do not contribute
anything proportional to  $\epsilon^{\mu \nu \rho \sigma}$.

\item Diagrams with charginos and neutralinos in the loop  yield
non-vanishing effects. CP non-conservation is introduced through phases
in the gaugino masses and the SUSY breaking parameter $\mu $.

\item For $V \equiv \gamma $ there is no CP-violating contribution.

\item For $V \equiv Z$ the analytical part of the calculation has been 
completed. The numerical evaluation remains to be done.

\item It appears that, barring unexpected cancellations,
contributions of $O(10^{-3})$ to $\tilde{\kappa}_Z$ and
$\tilde{\lambda}_Z$ are possible for light charginos
(masses $\sim 100 - 300$ GeV) and allowably large CP-violating phases.

\end{itemize}

\newpage

\item {\bf Perturbative top Yukawa coupling in SUSY for $\tan \beta <
1.6$} 

The top quark Yukawa coupling in the MSSM is given in standard
notation by 
\beqn
h_t(m_t) = m_t(m_t) \frac{\sqrt{2}}{v} \frac {\sqrt{1 + \tan ^2
\beta}}{\tan \beta}.
\eeqn
Here $v = (1/G_F\sqrt{2})^{1/2}$ and $\tan\beta$ is the ratio of two
Higgs VEVs -- the one yielding masses for up-type fermion over that
for down-types.
It is readily seen that $h_t$ {\em increases} as $\tan \beta$
decreases. Further, $h_t$ grows with the renormalisation scale $\mu $.
Requiring the perturbative limit $h_t(\mu) < \sqrt{4 \pi} $ to be
satisfied all the way upto $M_{GUT}$, one finds the bound $\tan \beta
> 1.6$. On the other hand, the $\tan \beta < 1.6 $ regime could be
phenomenologically interesting \cite{ksw}. Is it possible to evade the
$\tan \beta $ bound in models with intermediate mass scales? The
situation examined, as a part of this Working Group activity, is based
on the symmetry sequence $SU(4)_c \times SU(2)_L \times SU(2)_R
\stackrel {M_R}{\rightarrow } SU(3)_c
\times SU(2)_L \times U(1)_Y $.
By varying $M_R$ it is possible to find
which choice will ensure $h_t(\mu) < \sqrt{4 \pi} $ at all scales and
yield Yukawa couplings consistent with the known fermion masses.

\item {\bf $\theta_W$, fermion masses and horizontal $U(1)$ symmetry}

The quark and charged lepton masses satisfy the empirical relations:
\[
\frac{m_u}{m_t} = O(\lambda ^8); \;\; \frac{m_c}{m_t} = O(\lambda ^4);
\;\;
\frac{m_d}{m_b} = O(\lambda ^4);  
\]
\[
\frac{m_s}{m_b} = O(\lambda ^2); \;\; \frac{m_e}{m_{\tau}} = O(\lambda
^4); \;\; \frac{m_{\mu}}{m_{\tau}} = O(\lambda ^2).
\]
In addition, the following radiations are believed to hold at $M_{GUT}$.
\[
m_b = m_{\tau}, \;\; \;\; \frac{m_d m_s m_b}{m_e m_{\mu}
m_{\tau}} = O(1) 
\]
It has been shown that these relations can be reproduced in the MSSM by
introducing an additional $U(1)_H$ horizontal symmetry \cite{br}. The
$U(1)_H$ charges -- in other words, the Yukawa couplings -- are
appropriately chosen to reproduce the required mass matrix textures.
The procedure also entails the introduction of a electroweak gauge
singlet $U(1)_H$ nonsinglet field. It turns out that the charge
assignments are such that the $U(1)_H$ has anomalies.  If the anomalies
are cancelled by the Green-Schwarz mechanism ({\em i.e.}, if the model
originates from superstring theory), then that sets constraints on the
mixed {\em gauge} anomalies. This, in turn, relates the gauge coupling
constants, yielding  $\sin^2
\theta _W = \frac{3}{8}$, a result that also emerges from GUT models. 

As a part of the activities in this working group, a similar situation
was investigated in the context of R-parity violating SUSY. It is of
interest to examine whether this method might relate the ${\not \!
\!{R}}$-couplings $\lambda,
\; \lambda' \;\; {\rm and} \;\; \lambda''$ and might even forbid one or
more of them. This work is in progress.

\item {\bf Can gauge coupling unification and gaugino mass unification
be decoupled?}

The equality of the three gaugino masses, corresponding to the
$SU(3)$, $SU(2)$ and $U(1)$ gauge groups of the SM, is usually assumed
at the GUT scale as one of the boundary conditions on the
Renormalization Group evolution in the MSSM.  Can one construct a
consistent SUSY GUT in which this assumption is violated?  This
question has been addressed in the context of $SU(5)$ and a model was
constructed with nonunified gaugino masses following the mechanism
used in \cite{ekn}.  However, the model has a problem in that all
fermionic partners of scalars in an adjoint representation of $SU(5)$
are left massless at the GUT scale.  The issue of making a more
consistent model is to be explored further. 

\item {\bf Should squarks be degenerate?}

It is well-known that Flavour Changing Neutral Current (FCNC)
constraints following from the $K^0 - \overline{K}^0$, $B^0 -
\overline{B}^0$ and $D^0 - \overline{D}^0$ mass differences, $b
\rightarrow s \gamma $ {\em etc.}, set strong restrictions \cite{ns}
in SUSY 
models. SUSY contributions to FCNC {\em via} sparticle exchange are
usually tamed by choosing the squarks and sleptons to be almost mass
degenerate. A comparison of the modes of suppressing FCNC in the
supersymmetric extension of the SM can be summarised as:

\begin{center}
\begin{tabular}{ccc}

SM && SUSY \\
&& \\
&& \\
Quarks are degenerate      & $\&$ &     Squarks are degenerate \\
&& \\
\multicolumn{3}{c}{or}\\
&& \\
CKM matrix is diagonal     & $\&$ &     SCKM matrix is diagonal\\
\end{tabular}
\end{center}
A special case of the latter is when the CKM (SCKM) matrix is 
proportional to the identity matrix. Such a situation obtains if in
the SM the $u$-quark and $d$-quark 
mass matrices are matched while in SUSY it requires a matching of the
quark and squark mass matrices \cite{ns}.
The latter mode of satisfying the FCNC constraints in SUSY was examined
and it was concluded that (a) a more careful analysis is called for to
constrain the mass matrix structures and (b) the restrictions from
CP-violation could turn out to be significant.

\item{\bf Non-SUSY resolution of $R_b$ anomaly and the $Z-t-\bar{t}$
vertex} 

Technicolor models may provide a solution to the discrepancy between the
observed value of $R_b$ at LEP and its SM prediction.  Such models also
alter the $Z-t-\bar{t}$ vertex and may be
probed at the NLC. How big are these effects?
One of the models examined involves the symmetry breaking
$SU(N+2)_{ETC} \times SU(2)_L \times U(1)_Y \rightarrow SU(N)_{TC}
\times SU(2)_H \times SU(2)_L \times U(1)_Y \stackrel {u}{\rightarrow }
SU(N)_{TC} \times SU(2)_L \times U(1)_Y \stackrel {v}{\rightarrow }
SU(N)_{TC} \times U(1)_{EM}$.

The relevant Lagrangian can be written as:
\beqn
{\cal L}_{Z t \bar{t}} \sim \bar{t} (g_V \gamma^{\mu} F_V + g_A \gamma
^{\mu} \gamma _5 F_A) t Z_{\mu} 
\eeqn
where $g_{V,A}$ are the SM couplings and \cite{c}
\beqn
F_V = 1 - 5.15 (\frac{\xi ^2}{4} \frac{m_t}{4 \pi v} + \frac{s^4}{2 x})
, \;\;
F_A = 1 - 2.0 (\frac{\xi ^2}{4} \frac{m_t}{4 \pi v} + \frac{s^4}{2 x}).
\eeqn
Here $x = u^2/v^2 \gg 1$ is a measure of the relative magnitudes of
the heavy and light scales in the model and $\xi $ is a
model-dependent Clebsch Gordan factor. Also, $s \equiv \sin \phi$ where
$\phi $ is 
the neutral heavy boson -- light boson mixing angle. Two cases were
considered:

\begin{itemize}
\item Light case: Choosing typical values of $x = 20$ and $\xi = 1.4$
and fitting the $R_b$ data implies $\delta F_V \simeq -0.25, \; \;
\delta F_A \simeq -0.10$ and $\delta R_c/R_c \simeq -0.002$.

\item Heavy case: Using $x = 380$ and $\xi = 0.8$
and fitting the $R_b$ data implies $\delta F_V \simeq -0.05, \; \;
\delta F_A \simeq -0.02$ and $\delta R_c/R_c \simeq -0.002$.
\end{itemize}

Another model \cite{h}, that was also examined, has an additional $U(1)$
symmetry which couples to the third generation only and that too in the
following manner: (a) vectorially to the $\tau $, (b) axial vectorially
to the $t$ and (c) in a left-handed way to the $\nu _{\tau}$. All the
couplings have equal strengths. In this model $R_b$ receives a 
contribution through the mixing of the extra $U(1)$ gauge boson with
the $Z$ boson. The parameters of the model can be chosen to fit the
experimental value of $R_b$. This is found to imply $\delta F_V = 0$,
$\delta F_A \simeq -0.0275, \; \delta R_c/R_c = -0.008$. 
Recall that the $U(1)$ couples axial vectorially to the $t$-quark.

\item {\bf Maximal mixing and three degenerate neutrinos}

The maximal mixing between three generations of neutrinos may be
parametrised by the mixing matrix:
\beqn
U = \frac{1}{\sqrt{3}} \pmatrix{~ 1 & \omega ~ & \omega ^2 ~
\cr ~ 1 & \omega ^2 ~ & \omega ~
\cr ~ 1 & 1 ~ & 1 ~}.
\eeqn
This choice automatically satisfies the neutrinoless double beta decay
constraint since $<m> = m_0 \;\Sigma U_{e i}^2 \sim 0$. In this model
$P_{\nu_e \nu_e} = P_{\nu_e \nu_{\mu}} = P_{\nu_e \nu_{\tau}} = 1/3$.
This scenario was compared with the solar neutrino data and the
following results were obtained.
\[
\begin{array}{|c|c|c|} \hline
   &  Cl &  Ga  \\ \hline
{\rm Theory}: \;\; \phi({^8B}) \;\; {\rm from \;\; Kamioka} & 4.5 \pm
0.5 & 123^{+8}_{-6} \\ \hline
{\rm Experiment} & 2.78 \pm 0.35 & 75 \pm 9 \\ \hline
{\rm 2 \;\; flavour \;\; maximal \;\; mixing} & 3.64 \pm 0.4 &
65^{+7}_{-4} \\ \hline
{\rm 3 \;\; flavour \;\; maximal \;\; mixing} & 3.35 \pm 0.17 &
45.7^{+7}_{-3} \\ \hline
\end{array}
\]
Notice that in the three-flavour case, the
situation gets much worse for $Ga$, though there is a small
change towards the observed value for the $Cl$ experiment. 

\item {\bf Energy independent neutrino depletion and three generations}

Another alternative \cite{fv} -- in some sense complementary to the
previous one 
-- is the situation where the neutrino mass differences are such that
the neutrino oscillation probabilities are independent of the energy.
With three generations, the probabilities relevant to the solar
neutrino problem can then be written as:

$P_{\nu_e \nu_e} = 1 - \frac{1}{2} c_{13}^4 \sin ^2 2\theta_{12}
- \frac{1}{2} \sin ^2 2\theta_{13}$

\noindent and 

$P_{\nu_e \nu_{\mu}} + P_{\nu_e \nu_{\tau}} = \frac{1}{2} c_{13}^4 \sin
^2 2\theta_{12} + \frac{1}{2} \sin ^2 2\theta_{13}$

\begin{itemize}
\item It was found that the Kamiokande, $Cl$ and $Ga$ data cannot be
simultaneously  explained within this scenario at a 95\% CL. A small
allowed region in the parameter space is found at the $3 \sigma$ level.

\item If the flux of the $^8B$ neutrinos is taken from the Kamiokande
data, then a much better fit is obtained.
\end{itemize}

\item {\bf Field theoretic formulation of neutrino and 
charged lepton oscillations}

As a part of this project, several ideas in the recent literature
pertaining to oscillations were critically examined. The following
observations can be made.

\begin{itemize}
\item A careful field theoretic analysis of neutrino mixing leads to
the realization 
\cite{bv} that the vacuum corresponding to flavour basis states is
a coherent state obtained from a condensation of mass eigenstates. The
two are {\em unitarily nonequivalent}. This changes the momentum
dependence of the oscillation probability $P_{\nu_e \nu_e}$, but 
the usual form is recovered in the ultra-relativistic
limit (which is expected to be valid for neutrinos). Whether this idea
can be tested was investigated.

\item If the $\nu_{\mu}$ is a superposition of three mass eigenstates, 
then in pion decay ($\pi \rightarrow \mu + \nu_{\mu} $) one should
obtain three different momenta of the muon in consistency with energy
momentum conservation. This will affect the time evolution of the muon
\cite{ssw}. This idea was critically examined and it was concluded
that, for the neutrino masses usually considered, 
the idea cannot be tested in experiments. Questions were also raised
about the rigourousness of the result itself. 
\end{itemize}

\item {\bf Violation of Equivalence Principle, Gravitational effects
in neutrino physics}

In this project there were several talks and group discussions. 
Problems, that were identified for further pursuit, include:

\begin{itemize}

\item Constraints on the Violation of Equivalence Principle (VEP) from
existing data on neutrinos.

\item Analysis of the parameter space for three neutrino generations
including VEP and matter effects.

\item Gravitational oscillations of Ultra-High-Energy (UHE) neutrinos.

\item The issues of principle related to VEP.

\end{itemize}

\item {\bf Singlet neutrino mixing and primordial nucleosynthesis}

Sterile neutrinos $\nu_s$ are of much current interest and have been
invoked to 
explain neutrino oscillation data. If such a neutrino has a large mixing
with either $\nu_e$ or $\nu_{\mu}$ then sterile neutrinos will
equilibrate rather late in 
the early universe. This may contribute as much as 0.8 to $N_{\nu}$ --
the effective number of neutrinos. Such a large contribution would be
in conflict with the bounds from nucleosynthesis. Several possible
loopholes to this line of argument were considered  in this working
group.

\begin{itemize}
\item Most recent observational data on $^4He$ and $^2H$ abundances
allow $N_{\nu}$ upto 4.5 \cite{ss}.
\item Lepton and/or baryon number asymmetries are not usually taken
into consideration when bounds on $N_{\nu}$ are set from
nucleosynthesis.
\item Nonlinear $\nu_e - \nu_s$ feedback is important. 
\end{itemize}

\end{enumerate}

\vskip 30pt

\section {Acknowledgements:}

We thank all the participants in these two groups for their all-round
cooperation.  The work of AR has been supported by grants from the
Department of 
Science and Technology and the Council of Scientific and Industrial
Research, India.

\end{document}